\documentclass[twocolumn,english]{IEEEtran}
\usepackage[T1]{fontenc}
\usepackage[latin9]{inputenc}
\usepackage{float}
\usepackage{bm}
\usepackage{amsmath}
\usepackage{graphicx}
\usepackage{amssymb}

\makeatletter

\floatstyle{ruled}
\newfloat{algorithm}{tbp}{loa}
\floatname{algorithm}{Algorithm}

\usepackage{cite}
\usepackage{bm}
\usepackage{bbm}
\usepackage{mathrsfs}

\@ifundefined{showcaptionsetup}{}{%
 \PassOptionsToPackage{caption=false}{subfig}}
\usepackage{subfig}
\makeatother

\usepackage{babel}

\begin{document}

\title{SET: AN ALGORITHM FOR CONSISTENT MATRIX COMPLETION}

\author{Wei Dai and Olgica Milenkovic\\
 Department of Electrical and Computer Engineering, University
of Illinois at Urbana-Champaign\\
 Email: \{weidai07,milenkovic\}@illinois.edu}
\maketitle
\begin{abstract}
A new algorithm, termed subspace evolution and transfer (SET), is
proposed for solving the consistent matrix completion problem. In
this setting, one is given a subset of the entries of a low-rank matrix,
and asked to find \emph{one} low-rank matrix consistent with the given
observations. We show that this problem can be solved by searching
for a column space that matches the observations. The corresponding
algorithm consists of two parts --- subspace evolution and subspace
transfer. In the evolution part, we use a line search procedure to
refine the column space. However, line search is not guaranteed to
converge, as there may exist barriers along the search path that prevent
the algorithm from reaching a global optimum. To address this problem,
in the transfer part, we design mechanisms to detect barriers and
transfer the estimated column space from one side of the barrier to
the another. The SET algorithm exhibits excellent empirical performance
for very low-rank matrices. \end{abstract}
\thispagestyle{empty}

\pagestyle{empty}

\renewcommand{\thefootnote}{\fnsymbol{footnote}} \footnotetext[0]{The authors would like to thank Dayu Huang for his help in designing the employed line-search procedure, and to acknowledge useful discussions with Yoram Bresler, Justin Haldar, Ely Kerman, Angelia Nedich, and Zoi Rapti. Furthermore, the authors would also like to thank the authors of \cite{cai_singular_2008,lee_Bresler_admira:_2009,Haldar_Hernando_powerfactorization_2009,montanari_keshavan_matrix_2009} for providing online software packages for their matrix completion algorithms.} \renewcommand{\thefootnote}{\arabic{footnote}} \setcounter{footnote}{0}

\section{Introduction}

Suppose that we observe a subset of entries of a matrix. The matrix
completion problem asks when and how the matrix can be \emph{uniquely}
recovered based on the observed entries. This reconstruction task
is ill-posed and computationally intractable. However, if the data
matrix is known to have low-rank, exact recovery can be accomplished
in efficient manners, provided that sufficiently many entries are
revealed. Low-rank matrix completion problem has received considerable
interests due to its wide applications, see for example \cite{Candes_Recht_2008_matrix_completion}
for more details.

An efficient way to solve the completion problem is via convex relaxation.
Instead of looking at rank-restricted matrices, one can search for
the matrix with minimum nuclear norm, subject to data consistency
constraints. Although in general nuclear norm minimization is not
equivalent to rank minimization, the former approach recovers the
same solution as the latter if the data matrix satisfies certain incoherence
conditions \cite{candes_tao_power_2009}. More importantly, nuclear
norm minimization can be accomplished by polynomial complexity algorithms,
for example, semi-definite programming or singular value thresholding
(SVT) \cite{cai_singular_2008}.

There are other low-complexity alternatives. Based on the subspace
pursuit (SP) and CoSaMP algorithms for compressive sensing \cite{Dai_2009_Subspace_Pursuit,Tropp_2009_CoSaMP},
the authors of \cite{lee_Bresler_admira:_2009} developed the so called
ADMiRA algorithm. A modification of the power factorization algorithm
was used for matrix completion in \cite{Haldar_Hernando_powerfactorization_2009}.
Another approach for solving this problem, termed OptSpace, was described
in \cite{montanari_keshavan_matrix_2009}.

The problem considered in this paper and its algorithmic solution
differ from all previously published approaches. The problem at hand
is to identify \emph{one} low-rank matrix \emph{consistent} with the
observations. The solution may or may not be unique. In contrast,
most results in matrix completion deal with the somewhat more restrictive
requirement that the reconstruction \textit{\emph{is}} \textit{unique}.
Hence, our approach can be applied to scenarios where the matrix is
highly under-sampled, and where potentially many consistent solutions
exist. The relaxation on uniqueness allows for the empirically observed
performance improvement over other completion techniques.

To solve the consistent matrix completion problem, we propose an algorithm,
termed subspace evolution and transfer (SET). We show that the matrix
completion problem can be solved by searching for a column (or row)
space that matches the observations. As a result, optimization on
the Grassmann manifold, i.e., subspace evolution, plays a central
role in the algorithm. However, there may exist {}``barriers'' along
the search path that prevent subspace evolution from converging to
a global optimum. To address this problem, in the subspace transfer
part, we design mechanisms to detect and cross barriers. Empirical
simulations demonstrate the excellent performance of the proposed
algorithm.

Despite resembling the OptSpace algorithm \cite{montanari_keshavan_matrix_2009}
in terms of using optimization over Grassmann manifolds, our approach
substantially differs from this algorithm. Searching over only one
space (column or row space) represents one of the most significant
differences: in OptSpace, one searches \emph{both} column and row
spaces simultaneously, which introduces numerical and analytical difficulties.
Moreover, when optimizing over the column space, one has to take care
of {}``barriers'' that prevent the search procedure from converging
to a global optimum, an issue that was not addressed before since
it was obscured by simultaneous column and row space searches.

\vspace{-0.06in}

\section{Consistent Matrix Completion}

Let $\bm{X}\in\mathbb{R}^{m\times n}$ be an unknown matrix with rank
$r\ll\min\left(m,n\right)$, and let $\Omega\subset\left[m\right]\times\left[n\right]$
be the set of indices of the observed entries, where $\left[K\right]=\left\{ 1,2,\cdots,K\right\} $.
Define the projection operator $\mathfrak{P}_{\Omega}:\;\mathbb{R}^{m\times n}\rightarrow\mathbb{R}^{m\times n}$
by \begin{align*}
\bm{X} & \mapsto\bm{X}_{\Omega},\;\mbox{where }\left(\bm{X}_{\Omega}\right)_{i,j}=\begin{cases}
\bm{X}_{i,j} & \mbox{if }\left(i,j\right)\in\Omega\\
0 & \mbox{if }\left(i,j\right)\notin\Omega\end{cases}.\end{align*}
 The \emph{consistent matrix completion} problem is to find \emph{one}
rank-$r$ matrix $\bm{X}^{\prime}$ that is consistent with the observations
$\bm{X}_{\Omega}$, i.e., \begin{align}
\left(P0\right):\; & \mbox{find }\bm{X}^{\prime}\mbox{ such that }\nonumber \\
 & \mbox{rank}\left(\bm{X}^{\prime}\right)\le r\mbox{ and }\mathfrak{P}_{\Omega}\left(\bm{X}^{\prime}\right)=\mathfrak{P}_{\Omega}\left(\bm{X}\right)=\bm{X}_{\Omega}.\label{eq:P0}\end{align}
 This problem is well defined as $\bm{X}_{\Omega}$ is generated from
the matrix $\bm{X}$ with rank $r$ and therefore there must exist
\emph{at least one solution}. In this paper, like in other approaches
in \cite{lee_Bresler_admira:_2009,Haldar_Hernando_powerfactorization_2009,montanari_keshavan_matrix_2009},
we assume that the rank $r$ is given. In practice, one may try to
sequentially guess a rank bound until a satisfactory solution has
been found.

\section{The SET Algorithm}


\subsection{Why optimize over column spaces only?}

In this section, we show that the problem $\left(P0\right)$ is equivalent
to finding a column space consistent with the observations.

Let $\mathcal{U}_{m,r}$ be the set of $m\times r$ matrices with
$r$ orthonormal columns, i.e., $\mathcal{U}_{m,r}=\left\{ \bm{U}\in\mathbb{R}^{m\times r}:\;\bm{U}^{T}\bm{U}=\bm{I}_{r}\right\} .$
Define a function \begin{align}
f:\;\mathcal{U}_{m,r} & \rightarrow\mathbb{R}\nonumber \\
\bm{U} & \mapsto\underset{\bm{W}\in\mathbb{R}^{n\times r}}{\min}\left\Vert \bm{X}_{\Omega}-\mathfrak{P}_{\Omega}\left(\bm{U}\bm{W}^{T}\right)\right\Vert _{F}^{2},\label{eq:f_U}\end{align}
 where $\left\Vert \cdot\right\Vert _{F}$ denotes the Frobenius norm.
The function $f$ captures the consistency between the matrix $\bm{U}$
and the observations $\bm{X}_{\Omega}$: if $f\left(\bm{U}\right)=0$,
then there exists a matrix $\bm{W}$ such that the rank-$r$ matrix
$\bm{U}\bm{W}^{T}$ satisfies $\mathfrak{P}_{\Omega}\left(\bm{U}\bm{W}^{T}\right)=\bm{X}_{\Omega}$.
Hence, the consistent matrix completion problem is equivalent to \begin{align}
\left(P1\right):\; & \mbox{find }\bm{U}\in\mathcal{U}_{m,r}\;\mbox{such that }f\left(\bm{U}\right)=0.\label{eq:P1}\end{align}

The solution $f\left(\bm{U}\right)$ is not unique in the space $\mathcal{U}_{m,r}$.
An important property of $f$ is that $f\left(\bm{U}\right)=f\left(\bm{U}\bm{V}\right)$
for any $r$-by-$r$ orthogonal matrix $\bm{V}$, since $\bm{U}\bm{W}^{T}=\left(\bm{U}\bm{V}\right)\left(\bm{W}\bm{V}\right)^{T}$.
Hence, the function $f$ depends only on the subspace spanned by the
columns of $\bm{U}$, i.e., the $\mbox{span}\left(\bm{U}\right)$.
Note that all columns of the matrix of the form $\bm{U}\bm{W}^{T}$
lie in the linear subspace $\mbox{span}\left(\bm{U}\right)$. The
consistent matrix completion problem is essentially \emph{finding
a column space consistent with the observed entries}.

We find the following definitions useful for the exposition to follow.
The set of all $r$-dimensional linear subspaces in $\mathbb{R}^{n}$
is called the Grassmann manifold, and is denoted by $\mathcal{G}_{m,r}$.
Given a subspace $\mathscr{U}\in\mathcal{G}_{m,r}$, one can always
find a matrix $\bm{U}\in\mathcal{U}_{m,r}$, such that $\mathscr{U}=\mbox{span}\left(\bm{U}\right)$.
The matrix $\bm{U}$ is referred to as \emph{a} generator matrix of
$\mathscr{U}$. Although a given subspace $\mathscr{U}\in\mathcal{G}_{m,r}$
has multiple generator matrices, a given matrix $\bm{U}\in\mathcal{U}_{m,r}$
uniquely defines a subspace. For this reason, we henceforth use $\bm{U}$
to represent its generated subspace.

\subsection{The SET algorithm: a high level description}

Our algorithm aims to minimize the objective function $f\left(\bm{U}\right)$,
provided that the minimum value of $f\left(\bm{U}\right)$ is \emph{known
to be zero}. Ideally, a solution can be obtained by using a line search
procedure on the Grassmann manifold. Here, line search refers to iterative
refinements of the interval in which the function attains its minimum.
Hence, the {}``subspace evolution'' part of the algorithm reduces
to a well studied optimization method.

The main difficulty that arises during line search, and makes the
SET algorithm highly non-trivial is when during the search, one encounters
{}``barriers''. Careful inspection reveals that the objective function
$f$ can be decomposed into a sum of atomic functions, each of which
involves only one column of $\bm{X}_{\Omega}$ (see Section \ref{sub:Subspace-Transportation}
for details). Along the gradient descent path, these atomic functions
may not agree with each other: some decrease and some increase. Increases
of some atomic functions may result in {}``bumps'' in the $f$ curve,
which block the search procedure from global optima and are therefore
referred to as \emph{barriers}. The main component of the {}``transfer''
part of the algorithm is to identify whether there exist barriers
along the gradient descent path. Detecting barriers is in general
a very difficult task, since one does not know the locations of global
minima. Nevertheless, we observe that barriers can be detected by
the existence of atomic functions with inconsistent descent directions.
When such a scenario is encountered, the algorithm {}``transfers''
the starting point of line search to the other side of the barriers,
and proceeds from there. Such a transfer does not overshoot global
minima as we enforce consistency of the steep descent directions at
the points before and after the transfer.

In summary, we start with a randomly generated $\bm{U}\in\mathcal{U}_{m,r}$
and then refine it until $f\left(\bm{U}\right)=0$. At each iteration,
we first detect and then cross barriers if there are any, and then
perform line search. The details of subspace evolution and transfer
are given in Section \ref{sub:Subspace-Evolution} and \ref{sub:Subspace-Transportation}.
Simulation results are presented in Section \ref{sec:Simulations}.

\subsection{\label{sub:Subspace-Evolution}Subspace evolution}

Due to space limitation, we focus on the $r=1$ case in Sections \ref{sub:Subspace-Evolution}
and \ref{sub:Subspace-Transportation}. Furthermore, our exposition
aims to make the algorithmic details as transparent to the readers
as possible. The highly technical \emph{performance and complexity
analysis} of SET for \emph{both $r=1$ and $r>1$} is deferred to
the journal version of the paper.

For the optimization problem at hand, we shall refine the current
column space estimate $\bm{u}$ following the gradient descent direction.
Here, the lowercase letter $\bm{u}$ is used to emphasize that the
$\bm{U}$ matrix is a vector when $r=1$. Let $\bm{w}_{\bm{u}}$ be
a length-$n$ column vector that achieves $f\left(\bm{u}\right)$,
and let $\bm{X}_{r}=\bm{X}_{\Omega}-\mathfrak{P}_{\Omega}\left(\bm{u}\bm{w}_{\bm{u}}^{T}\right)$.
Then the gradient%
\footnote{The gradient is well defined almost everywhere in $\mathcal{U}_{m,r}$.%
} of $f$ at $\bm{u}$ is given by \begin{align}
\nabla_{\bm{u}}f & =-2\bm{X}_{r}\bm{w}_{\bm{u}}.\label{eq:gradient}\end{align}
 The gradient descent path is chosen to be \emph{the geodesic curve}
on the Grassmann manifold with direction $\bm{h}=-\nabla_{\bm{u}}f/\left\Vert \nabla_{\bm{u}}f\right\Vert _{F}$.
A geodesic surve is an analogue of a straight line in an Euclidean
space: given two points on the manifold, the geodesic curve connecting
them is the path of the shortest length on the manifold. According
to \cite[Theorem 2.3]{edelman_optimization_manifolds_1998}, the geodesic
curve starting from $\bm{u}$, along $\bm{h}$, is given by \begin{equation}
\bm{u}\left(t\right)=\bm{u}\cos t+\bm{h}\sin t,\quad t\in\left[0,\pi\right).\label{eq:geodesic-1d}\end{equation}
 We restrict $t$ to the interval $\left[0,\pi\right)$ because $f\left(\bm{u}\left(t\right)\right)$
has period $\pi$, i.e., $f\left(\bm{u}\left(t+\pi\right)\right)=f\left(-\bm{u}\left(t\right)\right)=f\left(\bm{u}\left(t\right)\right)$.
Interested readers are referred to \cite{edelman_optimization_manifolds_1998}
for more details on geodesics on the Grassmann manifold.

The subspace evolution part is designed to search for a minimizer
(in most cases, a local minimizer) of the function $f$ along the
geodesic curve. Our implementation includes two steps. The goal of
the first step is to identify an interval $\left[0,t_{\max}\right]$
that contains a minimizer. Since $f\left(t\right)$ is periodic, $t_{\max}$
is upper bounded by $\pi$. The second step is devoted to locating
the minimizer $t^{*}\in\left[0,t_{\max}\right]$ accurately by iteratively
applying the golden section rule \cite{gill_practical_optimization_1982}.
These two steps are described in Algorithm \ref{alg:SE}. The constants
are set to $\epsilon=10^{-9}$, $c_{1}=\left(\sqrt{5}-1\right)/2$,
$c_{2}=c_{1}/\left(1-c_{1}\right)$ and $itN=10$. Ideally, the starting
step size $\epsilon>0$ should be chosen as small as possible. We
fix it to a constant as computers only have finite precision and $10^{-9}$
is already sufficiently small in all our experiments.

\begin{algorithm}[tbh]
\caption{\label{alg:SE}Subspace evolution.}

\textbf{Input}: $\bm{X}_{\Omega}$, $\Omega$, $\bm{u}$, and $itN$.

\textbf{Output}: $t^{*}$ and $\bm{u}\left(t^{*}\right)$.

\textbf{Step A}: find $t_{\max}\le\pi$ such that $t^{*}\in\left[0,t_{\max}\right]$

Let $t^{\prime}=\epsilon\pi$. 
\begin{enumerate}
\item Let $t^{\prime\prime}=c_{2}\cdot t^{\prime}$. If $t^{\prime\prime}>\pi$,
then $t_{\max}=\pi$ and quit Step A. 
\item If $f\left(\bm{u}\left(t^{\prime\prime}\right)\right)>f\left(\bm{u}\left(t\right)\right)$,
then $t_{\max}=t^{\prime\prime}$ and quit Step A. 
\item Otherwise, $t^{\prime}=t^{\prime\prime}$. Go back to step 1). 
\end{enumerate}
\textbf{Step B}: numerically search for $t^{*}$ in $\left[0,t_{\max}\right]$.

Let $t_{1}=t_{\max}/c_{2}^{2}$, $t_{2}=t_{\max}/c_{2}$, $t_{4}=t_{\max}$,
and $t_{3}=t_{1}+c_{1}\left(t_{4}-t_{1}\right)$. Let $itn=1$. Perform
the following iterations. 
\begin{enumerate}
\item If $f\left(\bm{u}\left(t_{1}\right)\right)>f\left(\bm{u}\left(t_{2}\right)\right)>f\left(\bm{u}\left(t_{3}\right)\right)$,
then $t_{1}=t_{2}$, $t_{2}=t_{3}$, and $t_{3}=t_{1}+c_{1}\left(t_{4}-t_{1}\right)$. 
\item Else, $t_{4}=t_{3}$, $t_{3}=t_{2}$ and $t_{2}=t_{1}+\left(1-c_{1}\right)\left(t_{4}-t_{1}\right)$. 
\item $itn=itn+1$. If $itn>itN$, then quit the iterations. Otherwise,
go back to step 1). 
\end{enumerate}
Let $t^{*}=\underset{t\in\left\{ t_{1},\cdots,t_{4}\right\} }{\arg\min}f\left(\bm{u}\left(t\right)\right)$
and compute $\bm{u}\left(t^{*}\right)$. 
\end{algorithm}

\subsection{\label{sub:Subspace-Transportation}Subspace transfer}

Unfortunately, the objective function $f\left(\bm{u}\right)$ may
not be a convex function of $\bm{u}$. The described linear search
procedure may not converge to a global minimum because the search
path may be blocked by what we call {}``barriers''. We show next
how to overcome the problem introduced by barriers.

At this point, we formally introduce the decoupling principle: the
objection function $f\left(\bm{u}\left(t\right)\right)$ is the squared
Frobenius norm of the residue matrix; it can be decomposed as the
sum of the squared Frobenius norm of the residue columns. More precisely,
let $\bm{x}_{\Omega_{j}}\in\mathbb{R}^{m\times1}$ be the $j^{th}$
column of the matrix $\bm{X}_{\Omega}$. Let $\mathfrak{P}_{\Omega_{j}}:\;\mathbb{R}^{m\times1}\rightarrow\mathbb{R}^{m\times1}$
be the projection operator corresponding to the $j^{th}$ column,
defined by \begin{align*}
\bm{v} & \mapsto\bm{v}_{\Omega_{j}},\;\mbox{where}\;\left(\bm{v}_{\Omega_{j}}\right)_{i}=\begin{cases}
\bm{v}_{i} & \mbox{if }\left(i,j\right)\in\Omega\\
0 & \mbox{if }\left(i,j\right)\notin\Omega\end{cases}.\end{align*}
 Then the objective function $f\left(\bm{u}\left(t\right)\right)$
can be written as a sum of $n$ atomic functions: \begin{align}
f\left(\bm{u}\left(t\right)\right) & =\underset{\bm{w}\in\mathbb{R}^{n\times1}}{\min}\left\Vert \bm{X}_{\Omega}-\mathfrak{P}_{\Omega}\left(\bm{u}\bm{w}^{T}\right)\right\Vert _{F}^{2}\nonumber \\
 & =\sum_{j=1}^{n}\underbrace{\min_{\bm{w}_{j}\in\mathbb{R}}\left\Vert \bm{x}_{\Omega_{j}}-\mathfrak{P}_{\Omega_{j}}\left(\bm{u}\left(t\right)\bm{w}_{j}\right)\right\Vert _{F}^{2}}_{f_{j}\left(\bm{u}\left(t\right)\right)}.\label{eq:f-decoupling}\end{align}
 This principle is essential to understanding the behavior of $f\left(\bm{u}\left(t\right)\right)$.

\begin{figure}
\begin{centering}
\subfloat[\label{fig:contour}Contours of $f_{1}$.]{\begin{centering}
\includegraphics[scale=0.35]{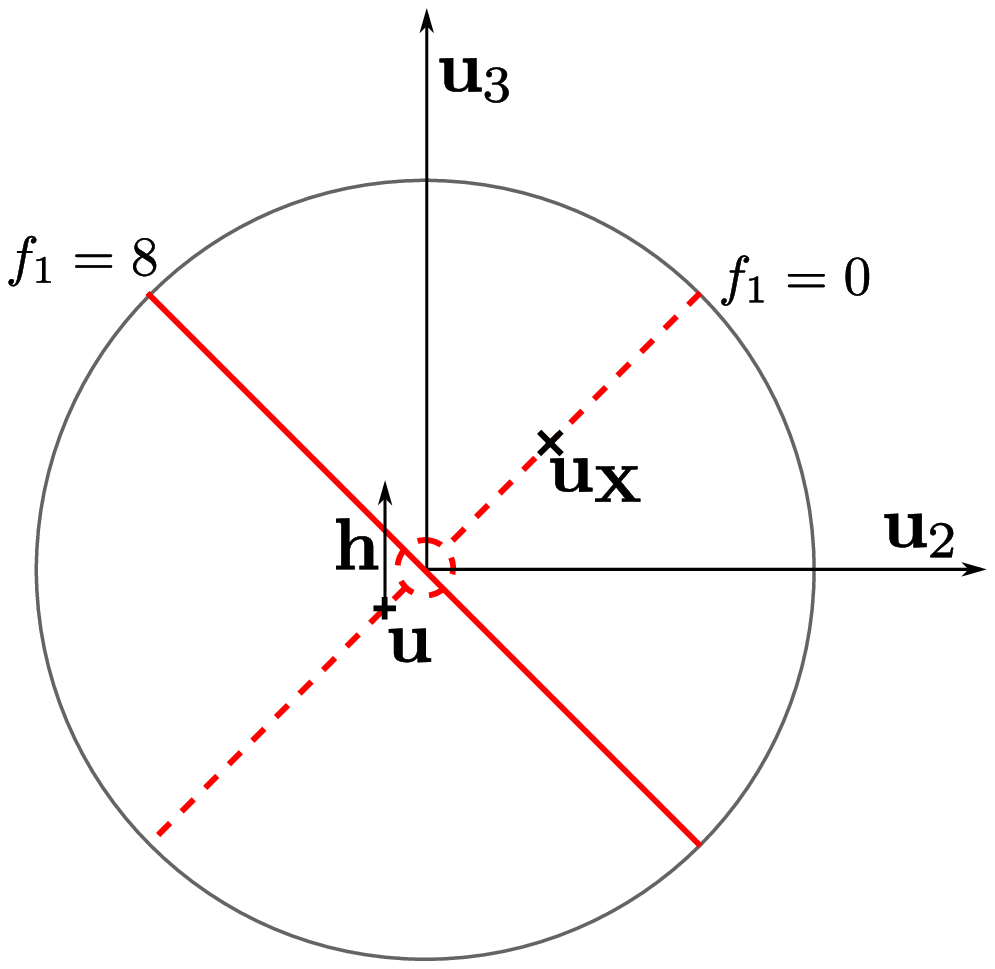}
\par\end{centering}

}\subfloat[\label{fig:search-path}Search paths with zooming in.]{\begin{centering}
\includegraphics[scale=0.35]{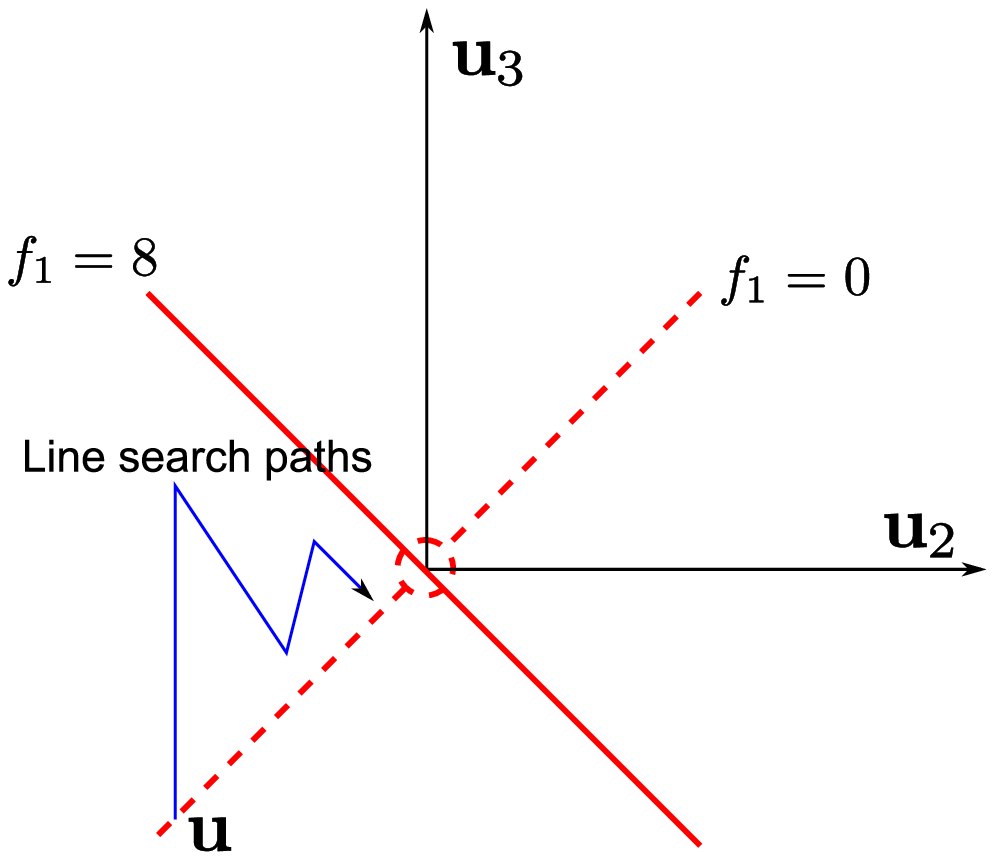}
\par\end{centering}

}
\par\end{centering}

\caption{\label{fig:example}An illustrative example for barriers.}

\end{figure}

The following example illustrates the concept of a barrier. Consider
an incomplete observation of a rank-one matrix $\left[\left[?,2,2\right]^{T},\left[2,?,1\right]^{T}\right],$
where question marks denote that the corresponding entries are unknown.
It is clear that the objective function $f\left(\bm{u}\left(t\right)\right)$
is minimized by $\bm{u}_{\bm{X}}=\frac{1}{\sqrt{6}}\left[2,1,1\right]^{T}$,
i.e., $f\left(\bm{u}_{\bm{X}}\right)=0$. Suppose that one starts
with the initial guess $\bm{u}=\frac{1}{\sqrt{102}}\left[-10,1,1\right]^{T}$.
The contours of the atomic function $f_{1}\left(\bm{u}\right)$, projected
on the plane spanned by $\bm{u}_{2}$ and $\bm{u}_{3}$, is depicted
in Fig. \ref{fig:contour}. All $\bm{u}$'s with $\bm{u}_{2}=-\bm{u}_{3}$
lie on the contour $f_{1}\left(\bm{u}\right)=8$. Computations show
that the gradient descent direction $\bm{h}$ is pointing upward.
However $f\left(\bm{u}\right)=f_{1}\left(\bm{u}\right)+f_{2}\left(\bm{u}\right)=0+f_{2}\left(\bm{u}\right)\le5<8$.
Any gradient descent algorithms can not pass through the contour $f_{1}\left(\bm{u}\right)=8$.
Careful tracking of several line search steps (Fig. \ref{fig:search-path})
shows that $\bm{u}\left(t\right)$ will approach $\left[-1,0,0\right]^{T}$,
but will never cross the contour $f_{1}=8$. That is, the contour
$f_{1}=8$ forms a {}``barrier'' for the line search procedure.

It is possible to detect barriers algorithmically. It can be verified
that $f_{j}\left(\bm{u}\left(t\right)\right)$ has a unique minimizer
and maximizer%
\footnote{The exception is that $\mathfrak{P}_{\Omega_{j}}\left(\bm{u}\right)$
and $\mathfrak{P}_{\Omega_{j}}\left(\bm{h}\right)$ are linearly dependent,
which happens with zero probability and is ignored here for simplicity.%
}, given by \begin{equation}
t_{\max,j}=\underset{t\in\left[0,\pi\right)}{\arg\max}f_{j}\left(\bm{u}\left(t\right)\right)\;\mbox{and}\; t_{\min,j}=\underset{t\in\left[0,\pi\right)}{\arg\min}f_{j}\left(\bm{u}\left(t\right)\right)\label{eq:toj-def}\end{equation}
 respectively. There are closed-form equations for these two quantities
given an initial vector $\bm{u}$ and a direction $\bm{h}$. We say
that the $k^{th}$ column of $\bm{X}_{\Omega}$ forms a barrier if
there exists a $j\in\left[n\right]$, $j\ne k$, such that 
\begin{enumerate}
\item the maximizer of $f_{k}$ appears before the minimizer of $f_{j}$,
i.e., $t_{\max,k}<t_{\min,j}<t_{\max,j}$; and 
\item the gradients of $f$ at $\bm{u}\left(0\right)$ and $\bm{u}\left(t_{\max,k}\right)$
are consistent (form a sharp angle), i.e., $\frac{d}{dt}f\left(\bm{u}\left(t\right)\right)|_{t=t_{\max,k}}<0$. 
\end{enumerate}
\vspace{0cm}

When a barrier is detected, we transfer $\bm{u}$ from one side of
it to the other. In our implementation, we focus on the closest barriers
to $\bm{u}$ to avoid overshooting. Define\[
\mathcal{J}=\left\{ j:\;\mbox{the }j^{th}\mbox{ column of }\bm{X}_{\Omega}\;\mbox{admits barriers}\right\} ,\]
 \begin{equation}
j^{*}=\underset{j\in\mathcal{J}}{\arg\min}\; t_{p,j},\;\mbox{and}\label{eq:def-j-star}\end{equation}
 \begin{align}
k^{*} & =\underset{k}{\arg\max}\;\left\{ t_{o,k}:\;\mbox{the }k^{th}\mbox{ column of }\bm{X}_{\Omega}\;\mbox{forms a barrier}\right.\nonumber \\
 & \qquad\qquad\qquad\left.\mbox{for the }j^{*^{th}}\mbox{ column of }\bm{X}_{\Omega}\right\} .\label{eq:def-k-star}\end{align}
 The subspace transfer part is described in Algorithm \ref{alg:ST}.

\begin{algorithm}[tbh]
\caption{\label{alg:ST}Subspace transfer}

\textbf{Input}: $\bm{X}_{\Omega}$, $\Omega$, and $\bm{u}$.

\textbf{Output}: $t_{st}$ and $\bm{u}\left(t_{st}\right)$.

\textbf{Steps}: 
\begin{enumerate}
\item Compute $t_{o,j}$ and $t_{p,j}$ for each column $j$ satisfying
$\mbox{rank}\left(\left[\bm{u}_{\Omega_{j}},\bm{h}_{\Omega_{j}}\right]\right)=2$. 
\item Suppose that there exist barriers.

\begin{enumerate}
\item find $j^{*}$ and $k^{*}$ according to (\ref{eq:def-j-star}) and
(\ref{eq:def-k-star}) respectively. 
\item Let $t_{st}=t_{o,k^{*}}$ and compute $\bm{u}\left(t_{st}\right)$. 
\end{enumerate}
\item Otherwise, $t_{st}=0$ and $\bm{u}\left(t_{st}\right)=\bm{u}$. 
\end{enumerate}

\end{algorithm}

\vspace{-0.1in}

\section{\label{sec:Simulations}Performance Evaluation}

Here, we introduce an error tolerance parameter $\epsilon_{e}>0$.
In practice, instead of requiring exact data matching, it usually
suffices to have $\left\Vert \mathfrak{P}_{\Omega}\left(\bm{X}^{\prime}\right)-\bm{X}_{\Omega}\right\Vert _{F}^{2}<\epsilon_{e}\left\Vert \bm{X}_{\Omega}\right\Vert _{F}^{2}$
for some small $\epsilon_{e}$. In our simulations, we set $\epsilon_{e}=10^{-6}$.

We tested the SET algorithm by randomly generating low-rank matrices
$\bm{X}$ and index sets $\Omega$. Specifically, we decompose the
matrix $\bm{X}$ into $\bm{X}=\bm{U}_{\bm{X}}\bm{S}_{\bm{X}}\bm{V}_{\bm{X}}^{T}$,
where $\bm{U}_{\bm{X}}\in\mathcal{U}_{m,r}$, $\bm{V}_{\bm{X}}\in\mathcal{U}_{n,r}$,
and $\bm{S}_{\bm{X}}\in\mathbb{R}^{r\times r}$. We generate $\bm{U}_{\bm{X}}$
and $\bm{V}_{\bm{X}}$ from the isotropic distribution on the set
$\mathcal{U}_{m,r}$ and $\mathcal{U}_{n,r}$, respectively. The entries
of the $\bm{S}_{\bm{X}}$ matrix are independently drawn from the
standard Gaussian distribution $\mathcal{N}\left(0,1\right)$. This
step is important in order to guarantee the randomness in the singular
values of $\bm{X}$. The index set $\Omega$ is randomly generated
from the uniform distribution over the set $\left\{ \Omega^{\prime}\subset\left[m\right]\times\left[n\right]:\;\left|\Omega^{\prime}\right|=\left|\Omega\right|\right\} $.

\begin{figure}[tbh]
\begin{centering}
\includegraphics[scale=0.45]{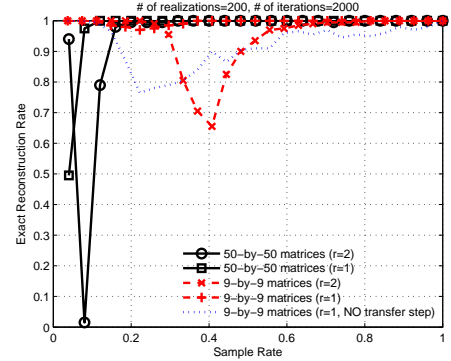} 
\par\end{centering}

\caption{\label{fig:SET-performance}Performance of the SET algorithm.}

\end{figure}

\begin{figure}[tbh]
\begin{centering}
\includegraphics[scale=0.45]{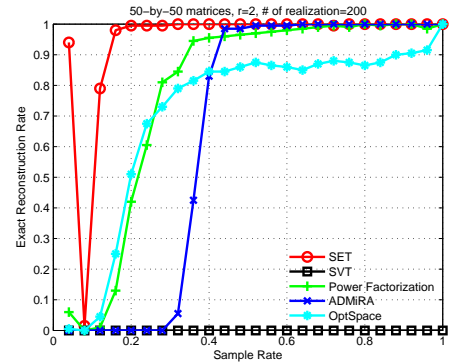} 
\par\end{centering}

\caption{\label{fig:SET-performance-comparison}Performance comparison.}

\end{figure}

The performance of the SET algorithm is excellent. We tested different
matrices with different ranks and different sampling rates, defined
as $\left|\Omega\right|/\left(m\times n\right)$. The performance
is shown in Fig. \ref{fig:SET-performance}. The performance improvement
due to the transfer step is significant. We also compare the SET algorithm
to other matrix completion algorithms%
\footnote{Though the SVT algorithm is not designed to solve the problem (P0),
we include it for completeness. In the standard SVT algorithm, there
is no explicit constraint on the rank of the \emph{reconstructed}
matrix. For fair comparison, we take the best rank-$r$ approximation
of the reconstructed matrix, and check whether it satisfies the performance
criterion.%
}. As shown in Figure \ref{fig:SET-performance-comparison}, the SET
algorithm outperforms all other tested completion approaches. For
most realizations, the SET algorithm needs less than 500 iterations
to converge. However, there are examples for which the reconstruction
error is still large after 2000 iterations. Studying such realizations
indicates that the major reason for this phenomena is a slow convergence
rate: after many iterations the barriers are still too far away from
space $\bm{U}$ to be detectable. One future research direction is
therefore to speed up the SET algorithm. As a final remark, we notice
that there exist a critical range of sampling rates, in which the
performance deteriorates. As can be observed from the figures, this
range shifts to the right as the rank increases.

\bibliographystyle{ieeetr}
\bibliography{MatrixCompletion}

\end{document}